\documentclass{article}
\usepackage{spconf,amsmath,graphicx,hyperref}
\usepackage{algorithm}
\usepackage{booktabs}
\usepackage{algpseudocode}

\title{Image Reconstruction from Phase with Untrained Neural Priors}
\name{Ene Meco, Ahmet Enis Cetin\thanks{© 2027 IEEE. Personal use of this material is permitted. Permission from IEEE must be obtained for all other uses.}}
\address{Dept. of Electrical and Computer Engineering, University of Illinois at Chicago, IL, 60607, USA}
\begin{document}
%
\maketitle

\begin{abstract}
Fourier phase encodes important spatial image structure, but recovering an image without measured spectral magnitude requires additional constraints and leaves absolute intensity ambiguous. We propose a projection-based two-stage framework that combines Fourier-phase and spatial-support constraints with an image-specific neural prior. The first stage alternates constraint enforcement with regularized neural-prior updates, while the second performs phase/support refinement alone with guaranteed convergence. We evaluate two neural-prior implementations on the same 77 microscopy images and compare them with a constraint-only baseline. After 500 final refinement passes, the best-performing variant achieves 31.41~dB pooled PSNR, 35.75~dB mean PSNR, and 0.9531 mean SSIM, improving pooled PSNR by 1.51~dB and reducing pooled MSE by 29.3\% relative to the baseline. The results demonstrate the benefit of combining neural guidance with explicit constraint refinement at the evaluated iteration budget, while showing that lower phase residual alone does not guarantee greater reconstruction accuracy.
\end{abstract}
\section{Introduction}
\label{sec:intro}

Fourier phase carries substantial spatial image information,
but does not determine absolute intensity~\cite{oppenheim1981}.
Classical studies established reconstruction and identifiability
results under additional assumptions, including known spatial
support~\cite{hayes1980,levi1983}. A single inverse Fourier
transform with unit magnitude generally remains inadequate,
motivating iterative reconstruction using phase and support
constraints.

Image-specific neural priors provide a complementary source of
regularization. Deep image prior optimizes a randomly initialized
network using a single observation~\cite{ulyanov2020}.
SelfDeblur extends this approach through joint image and kernel
estimation~\cite{ren2020,selfdeblurcode}, while Self-Diffusion
learns a denoiser through optimization under scheduled
noise~\cite{NEURIPS2025_7045440a}. Neither requires dataset
pretraining, making them suitable candidates for guiding
phase-only reconstruction.

We propose a two-stage framework that combines phase and support
constraints with neural-prior updates, followed by phase/support
refinement alone. Experiments on 77 microscopy images examine
the contributions of both stages against a constraint-only
baseline. The results show that an appropriate neural prior
can improve reconstruction quality at the evaluated iteration
budget, while lower phase residual alone does not guarantee
greater image-domain accuracy.

\vspace{-0.25cm}
\section{Related Work}
\label{sec:related}
\vspace{-0.25cm}
Phase-only reconstruction has a long history in signal processing.
Oppenheim and Lim emphasized the structural information carried by
phase~\cite{oppenheim1981}; Hayes \emph{et al.} studied reconstruction
conditions~\cite{hayes1980}, and Levi and Stark considered iterative
restoration using constraint sets~\cite{levi1983}. The set of signals or images with a known phase $\phi_d (\omega_1, \omega_2)$ constitute a closed and convex set in $\ell_2$ or $\ell_2$x$\ell_2$ \cite{youla1982image, sezan1982image,ccetin1988convolution, cetin1987procedure}:
\begin{equation}
C_\phi = \{ x \ | \ \phi_x (\omega_1, \omega_2) = \phi_d (\omega_1, \omega_2) \ \} 
\end{equation}
where $\phi_x (\omega_1, \omega_2)$ is the phase of the Fourier Transform $X(\omega_1, \omega_2) = |X(\omega_1, \omega_2)| e^{j \phi (\omega_1, \omega_2)} $ of the image $x$. We assume that $\phi_d (\omega_1, \omega_2)$ is known. The set of images with a known support region is also a closed and convex set. 
\begin{equation}
C_s = \{ x \ | \ x (n_1, n_2) = 0, \ \ (n_1, n_2) \not\in I \ \} 
\end{equation}
where $I$ is the known image support region in $Z^2$. Similarly, positive valued signals are closed and convex in l2xl2. Therefore the iterates obtained by successive orthogonal projections onto the sets (POCS) $C_\phi$ and $C_s$ defined as follows 
\begin{equation}
x_{k+1}= P_\phi ( P_s (x_k )),  \ \ k=0,1,2,...
\end{equation}
always converges for all initial conditions $x_o$. In Eq. 3, $P_\phi$ and $P_s$ represent the orthogonal projection operations onto the sets  $C_\phi$ and $C_s$, respectively. The intersection set $C_\phi \cap C_s$ is very large because $x$ and its scaled version $ax$ has the same phase.
The phase-only recovery problem is also formulated through convex optimization
~\cite{wu2016}, while more recent work studies identifiability up to
positive scale~\cite{chen2023}.

DIP~\cite{ulyanov2020} and SelfDeblur~\cite{ren2020} provide
image-specific neural priors without offline training. In our setting,
however, the observation is Fourier phase rather than a spatially
blurred image. The neural block therefore acts as an intermediate
image prior inside an explicit phase/support reconstruction loop,
rather than as a standalone deblurring method. We additionally
evaluate Self-Diffusion as an alternative untrained prior under the
same reconstruction constraints.

Projection-based signal and image recovery provides an important
foundation for the present approach. Youla and Webb developed the
method of projections onto convex sets for image restoration
~\cite{youla1982image}, with subsequent numerical applications by
Sezan and Stark~\cite{sezan1982image}. Levi and Stark later treated
restoration from Fourier phase and magnitude using generalized
projections~\cite{levi1987restoration}. Related work by
\c{C}etin et al. assumes zero-phase for FIR filter design \cite{cetin1997equiripple} and introduced a convolution-based signal-recovery
framework in which Fourier-domain magnitude and phase constraints can
be incorporated~\cite{ccetin1988convolution}.

\section{Iterative Reconstruction Method}
\label{sec:method}

\subsection{Successive Projections onto Convex Sets}

Let $x_\star\in[0,1]^{M\times N}$ denote a real reference image.
Operator $E$ embeds it in the top-left corner of a zero-filled
$2M\times2N$ canvas, $P_I$ imposes that support, and $F$ denotes
the unnormalized two-dimensional DFT on the doubled canvas and $F^{-1}$ is the IDFT.
The inverse transform includes the normalization factor $1/(4MN)$.

The measured Fourier phase is
\begin{equation}
\phi_d(\omega)
=
\arg\!\left((F[Ex_\star])(\omega)\right),
\label{eq:measurement}
\end{equation}
where $\omega$ denotes the  two-dimensional (2D) frequency on the 2D
Fourier grid.
No reference magnitude is supplied.
At zero Fourier coefficients, the measurement simulator uses
the convention $\arg(0)=0$.

Define the phase-replacement operator by
\begin{equation}
R_\phi(|Y|)(\omega)
=
|Y(\omega)|\exp\!\left(j\phi_d(\omega)\right).
\label{eq:phase_replacement}
\end{equation}
This operation retains the current Fourier magnitude and imposes
the measured phase. The operator $R_\phi$ performs phase replacement and is
the orthogonal projection onto the prescribed phase set $C_\phi$.
The corresponding update equation is
\begin{equation}
v= T(x)=
P_I\!\left(
\!\left[
F^{-1}\!\left[R_\phi(F[Ex])\right]
\right]
\right).
\label{eq:operators}
\end{equation}
where $P_I $ is the projection onto the set $C_I$, which zeros out all the nonzero terms outside the MxN support region, $I$. During the implementation of one iteration cycle we 
perform zero-padding to a 2Mx2N region for the FFT computation, phase replacement, inverse
transformation, and top-left support cropping.
Nonnegativity is also an orthogonal projection in $\ell_2$ and it can be imposed during the same iteration cycle as shown in Fig. 1. 

Because $x_\star$ and $cx_\star$, $c>0$, have identical Fourier
phase, absolute intensity cannot be recovered from $\phi_d$ alone.
We use the fixed initialization
\begin{equation}
u^0=
\operatorname{ReLU}\!\left(
P_I\!\left(
\operatorname{Re}\!\left[
F^{-1}\!\left[\exp(j\phi_d)\right]
\right]
\right)
\right),
\qquad
x^0=a u^0,
\label{eq:initial}
\end{equation}
where the exponential is applied elementwise,
$\operatorname{ReLU}(v)=\max(v,0)$, and 
$a$ 
is a constant.

\begin{figure}[t]
    \centering
    \includegraphics[width=0.6\columnwidth]{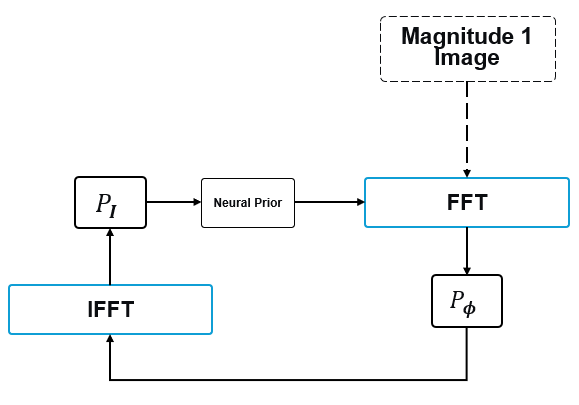}
    \caption{Flow diagram of the phase-only reconstruction
    framework. Starting from the constant magnitude initialization,
    each iteration applies Fourier-phase replacement, inverse
    transformation, support cropping, and nonnegativity enforcement.
    The constrained image is processed by the neural prior and
    combined with its output through relaxation.}
    \label{fig:framework}
\end{figure}
\vspace{-0.25cm}
\subsection{Neural-prior stage}
\vspace{-0.25cm}
At the $k$-th iteration, phase replacement is followed by support and nonnegativity enforcement:
\begin{equation}
\vspace{-0.25cm}
z^k=\operatorname{ReLU}\!\left(T(x^{k-1})\right).
\label{eq:z}
\end{equation}
where the operator $T$ represents an iteration cycle in Eq. (6).
The constrained image $z^k$ is then supplied to an untrained neural
block, which returns an $M\times N$ reconstruction $r^k$.

For both SelfDeblur and Self-Diffusion, network weights are randomly initialized for each image and optimized using its phase-derived observations, without loading pretrained weights or transferring learned parameters between images.

We introduced neural operators into the loop to regularize the reconstruction process similar to the plug and play framework \cite{kamilov2023plug}.
For the deep neural network SelfDeblur [5], we use the official implementation's image
generator $g_\theta$ and normalized kernel generator
$h_\psi$~\cite{selfdeblurcode}.
With $\mathcal B_h(g)$ denoting valid cross-correlation,
write the predicted fitting observation as
\begin{equation}
\widehat z_{k,t}
=
\mathcal B_{h_\psi}(g_\theta).
\label{eq:fitting_prediction}
\end{equation}
The fitting loss is
\begin{equation}
\ell_{k,t}=
\begin{cases}
\dfrac{\|\widehat z_{k,t}-z^k\|_F^2}{MN},
    & t\leq1000,\\[4pt]
1-\operatorname{SSIM}_{\rm fit}
    (\widehat z_{k,t},z^k),
    & t>1000,
\end{cases}
\label{eq:loss}
\end{equation}
where $t$ counts cumulative optimizer updates for the current
image across outer iterations.
For SelfDeblur, network parameters, optimizer state, and base
latent inputs persist across iterations but are initialized
anew for each image.

We also studied another network, Self-Diffusion~\cite{NEURIPS2025_7045440a} which provides an alternative
neural block based on optimizing an untrained denoiser under
scheduled noise.
It operates on the same phase-derived input $z^k$ and generates the next iterate.

For either neural prior, the next estimate is
\begin{equation}
\vspace{-0.25cm}
x^k=(1-\rho)z^k+\rho r^k,
\qquad 0<\rho\leq1.
\label{eq:relax}
\end{equation}
The fitting target $z^k$ remains fixed during each inner optimization
block; no gradients propagate through preceding outer iterations.
\vspace{-0.25cm}

\subsection{Proof of Convergence}
\vspace{-0.25cm}
After $K$ iterations, the neural block is removed. Setting
$v^0=x^K$, we only apply
\begin{equation}
\vspace{-0.25cm}
v^\ell=T(v^{\ell-1}),\qquad \ell=1,\ldots,L.
\vspace{-0.1cm}
\label{eq:tail}
\end{equation}
These passes retain signed values and impose no clipping to $[0,1]$. The orthogonal projection operator
 $R_\phi$ is nonexpansive:
\begin{equation}
\vspace{-0.25cm}
\|R_\phi(A)-R_\phi(B)\|_F
=\||A|-|B|\|_F\leq\|A-B\|_F.
\label{eq:nonexpansive}
\end{equation}
Since padding is isometric and cropping and real-part extraction are
contractions, $T$ is also nonexpansive. For exact phase,
$T(cx_\star)=cx_\star$ for $c>0$, hence
\begin{equation}
\vspace{-0.20cm}
\|v^\ell-cx_\star\|_F
\leq
\|v^{\ell-1}-cx_\star\|_F.
\label{eq:error}
\vspace{-0.20cm}
\end{equation}
Thus the final exact-phase passes cannot increase error to a correctly
scaled, and this establishes
convergence of the complete procedure to the intersection of the sets $C_I$ and $C_\phi$. So, Algorithm 1 is globally convergent \cite{bregman1967relaxation},\cite{youla1982image}. The iterations with neural networks in the loop provide an initial estimate to the globally convergent POCS methods.
\begin{algorithm}[t]
\caption{Phase-only reconstruction with a neural prior}
\label{alg:method}
\begin{algorithmic}[1]
\Require Phase $\phi$, support $(M,N)$, $a,\rho,K,L$
\State Initialize $x^0$ using \eqref{eq:initial}
\State Initialize neural-prior state
\For{$k=1$ to $K$}
    \State $z^k\gets\operatorname{ReLU}(T(x^{k-1}))$
    \State Fit/apply neural prior to $z^k$; obtain $r^k$
    \State $x^k\gets(1-\rho)z^k+\rho r^k$
\EndFor
\State $v^0\gets x^K$
\For{$\ell=1$ to $L$}
    \State $v^\ell\gets T(v^{\ell-1})$
\EndFor
\State \Return $v^L$
\end{algorithmic}
\end{algorithm}

\begin{figure}[t]
    \centering
    \includegraphics[width=\columnwidth]
    {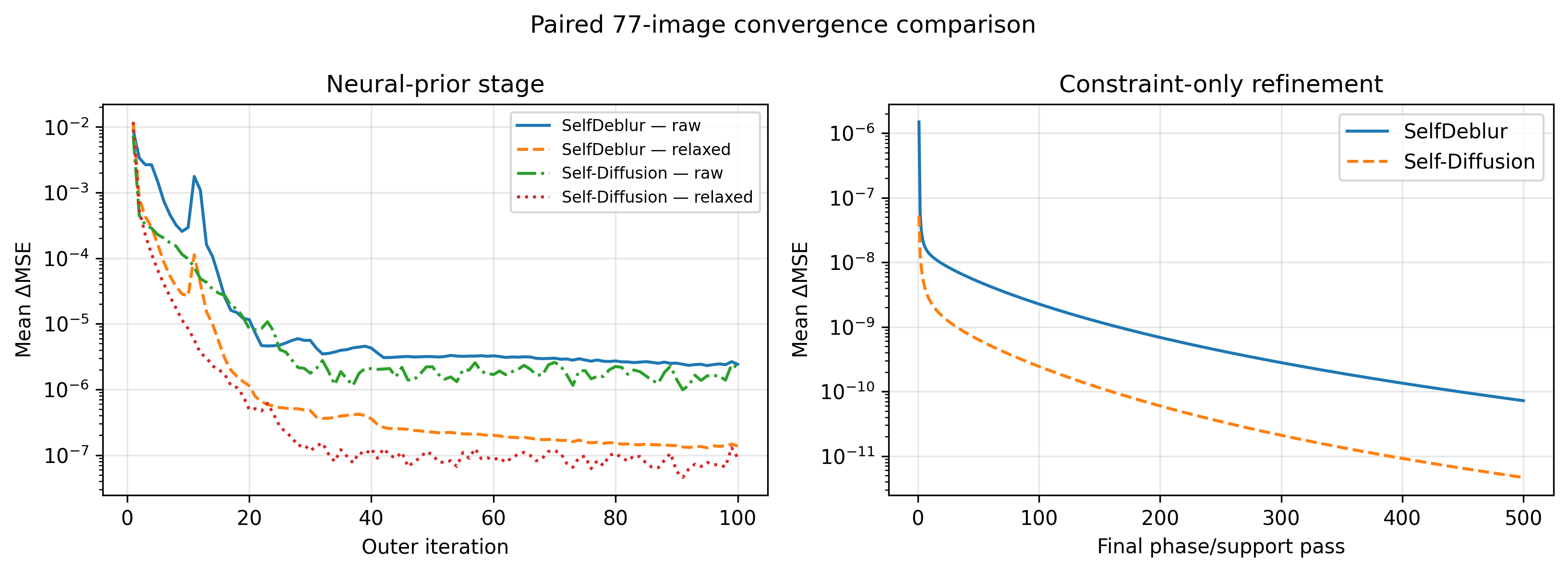}
    \caption{Convergence diagnostics on the paired 77-image subset.
    Left: mean squared change between successive neural-prior outputs
    and relaxed estimates during the outer stage. Right: mean squared
    change between successive estimates during the final
    phase/support-only refinement.}
    \label{fig:convergence}
\end{figure}

\begin{table*}[!t]
\centering
\caption{Paired comparison on the same 77 images. Mean PSNR and SSIM
are reported as mean $\pm$ standard deviation across images.}
\label{tab:prior_comparison_full}
\scriptsize
\setlength{\tabcolsep}{3.5pt}
\begin{tabular}{llcccccc}
\toprule
Method & $L$
& Pooled MSE
& Pooled PSNR
& Mean PSNR
& Mean SSIM
& Phase residual\\
& & & (dB) & (dB) & & & \\
\midrule

Phase/support & 100
& $1.785{\times}10^{-3}$
& 27.48
& $28.21 \pm 2.44$
& $0.6973 \pm 0.0892$
& $1.79{\times}10^{-2}$\\

Phase/support & 500
& $1.080{\times}10^{-3}$
& 29.66
& $32.17 \pm 5.56$
& $0.8564 \pm 0.1010$
& $6.17{\times}10^{-3}$\\

\midrule

Self-Diff. & 0
& $1.132{\times}10^{-3}$
& 29.46
& $33.78 \pm 7.17$
& $0.9392 \pm 0.0583$
& $6.85{\times}10^{-3}$\\

Self-Diff. & 300
& $1.101{\times}10^{-3}$
& 29.58
& $35.34 \pm 10.04$
& $0.9408 \pm 0.0628$
& $7.30{\times}10^{-4}$\\

Self-Diff. & 500
& $1.099{\times}10^{-3}$
& 29.59
& $35.59 \pm 10.66$
& $0.9408 \pm 0.0631$
& $3.62{\times}10^{-4}$\\

\midrule

SelfDeblur & 0
& $9.86{\times}10^{-4}$
& 30.06
& $31.71 \pm 3.70$
& $0.8925 \pm 0.0656$
& $1.48{\times}10^{-2}$\\

SelfDeblur & 300
& $7.44{\times}10^{-4}$
& 31.28
& $35.04 \pm 6.47$
& $0.9467 \pm 0.0644$
& $2.85{\times}10^{-3}$\\

SelfDeblur & 500
& $\mathbf{7.24{\times}10^{-4}}$
& \textbf{31.41}
& $\mathbf{35.75 \pm 7.35}$
& $\mathbf{0.9531 \pm 0.0632}$
& $1.53{\times}10^{-3}$\\

\bottomrule
\end{tabular}
\end{table*}

\begin{figure}[t]
    \centering
    \includegraphics[width=\columnwidth]
    {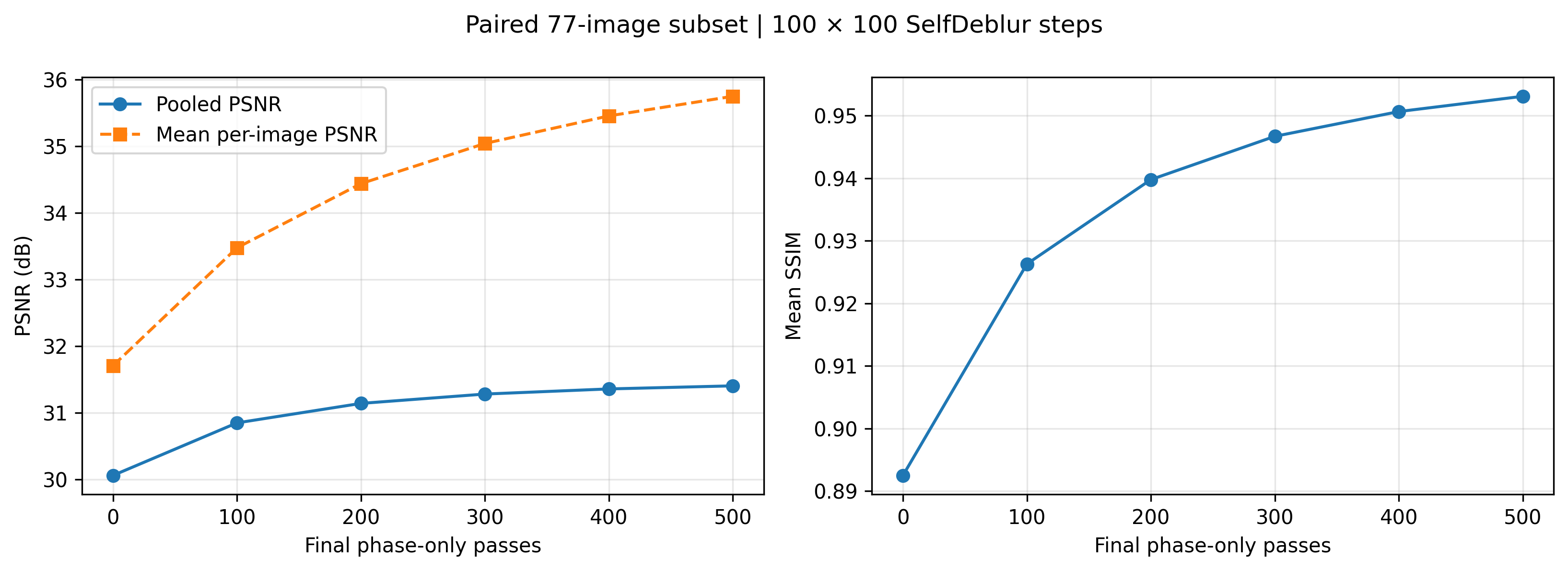}
    \caption{Effect of the final phase/support refinement on
    SelfDeblur over the paired 77-image subset. Left: pooled
    and mean per-image PSNR. Right: mean SSIM.}
    \label{fig:quality_vs_passes}
\end{figure}

\vspace{-0.25cm}
\begin{figure}[t]
    \centering
    \includegraphics[width=\columnwidth]{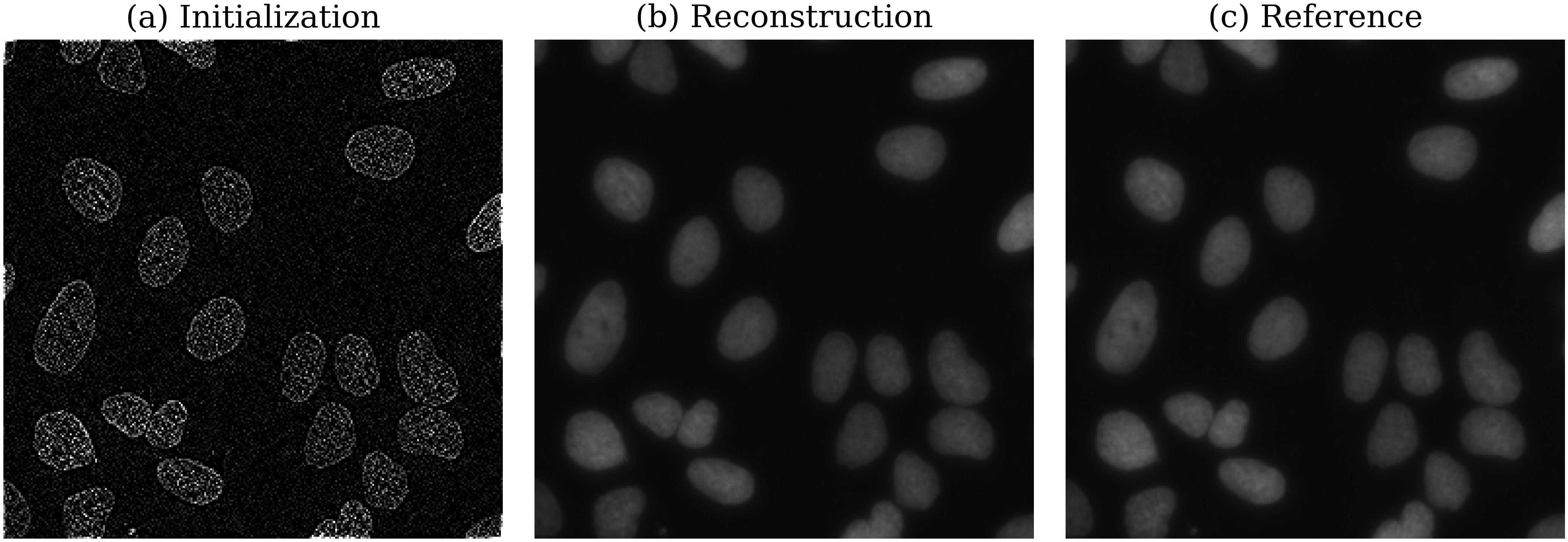}
    \caption{Phase-only reconstruction of a BBBC006 image:
    (a) initialization scaled by $a=50$;
    (b) SelfDeblur reconstruction after 100 outer iterations and
    500 final phase/support passes;
    (c) reference. All panels share the display range $[0,1]$.}
    \label{fig:qualitative_reconstruction}
\end{figure}
\vspace{-0.25cm}
\section{Experimental Protocol}
\label{sec:experiments}
\vspace{-0.25cm}
BBBC006v1 microscopy images~\cite{ljosa2012,bbbc006} are evaluated
using channel w1, focal plane $z=16$, and one deterministic
$256\times256$ center crop per field of view. Integer intensities are
divided by 4095 with no image-specific contrast normalization. Each
crop generates a $512\times512$ exact Fourier-phase observation; no
measurement noise is added.

For the paired experiment, 77 image identifiers were randomly selected
with a fixed seed from the samples successfully reconstructed by both
SelfDeblur and Self-Diffusion. The same 77 images are used for both
methods and correspond to approximately 10\% of the 768-image dataset.
All settings are fixed across images with no reference-based stopping
or image-specific model selection.

The reconstruction hyperparameters, including the initialization
amplitude $a=50$ and relaxation $\rho=0.25$, were selected through
ablation studies. The selected settings are held fixed across the
77-image comparison, without image-specific parameter adjustment.


Both methods use $K=100$ outer phase/support iterations.
At each outer iteration, SelfDeblur performs
$J_{\mathrm{SB}}=100$ joint image/kernel optimization
iterations~\cite{selfdeblurcode}.
Self-Diffusion uses $S=20$ noise steps, with
$J_{\mathrm{SDI}}=5$ optimization iterations at each noise
step~\cite{NEURIPS2025_7045440a}.
Its inner budget is therefore
$S J_{\mathrm{SDI}}=20\times5=100$ optimization iterations
per outer iteration.
Consequently, both methods perform $10{,}000$ inner training
iterations per image before the final phase/support refinement.
This matches the number of optimization iterations; the
computational cost of an iteration depends on the neural
architecture and fitting procedure.
Runtime is therefore reported separately.

For SelfDeblur, the auxiliary kernel has size $21\times21$,
the random seed is zero, and the initial image/kernel Adam
learning rates are $10^{-2}$ and $10^{-4}$, respectively.
Network and optimizer states persist across outer iterations
within each image.
For both methods, final refinement is recorded at
$L\in\{0,100,200,300,400,500\}$.

Reconstruction quality is measured using MSE, PSNR, and SSIM with
fixed data range one:
\begin{equation}
\mathrm{MSE}=\frac{\|v-x_\star\|_F^2}{MN},\qquad
\mathrm{PSNR}=-10\log_{10}\mathrm{MSE}.
\label{eq:metrics}
\vspace{-0.25cm}
\end{equation}
SSIM uses an $11\times11$ Gaussian window with standard deviation
1.5. No clipping, alignment, or reference-fitted intensity rescaling
is used. We report pooled PSNR, mean per-image PSNR, mean SSIM,
runtime, and the relative phase residual
\begin{equation}
\vspace{-0.3cm}
e_\phi(v)=
\frac{
\left\|F[Ev]-|F[Ev]|\odot\exp(j\phi_d)\right\|_F
}{
\max\!\left(\|F[Ev]\|_F,10^{-12}\right)
}.
\label{eq:residual}
\end{equation}
\vspace{-0.25cm}
\section{Results and Discussion}
\label{sec:results}
\vspace{-0.25cm}
Table~\ref{tab:prior_comparison_full} reports the paired comparison on the
same 77 images. Before the final constraint-only refinement
($L=0$), Self-Diffusion achieves higher mean PSNR and SSIM than
SelfDeblur, with 33.78~dB and 0.9392 versus 31.71~dB and 0.8925.
It also has the lower phase residual, although SelfDeblur already has
the higher pooled PSNR, 30.06 versus 29.46~dB.

The final phase/support stage affects the two priors differently.
For SelfDeblur, Fig.~\ref{fig:quality_vs_passes} shows a clear
improvement as additional refinement passes are applied: pooled PSNR
increases to 31.41~dB, mean PSNR to 35.75~dB, and mean SSIM to
0.9531 after 500 passes. In contrast, Self-Diffusion changes only
slightly in pooled PSNR and SSIM, reaching 29.59~dB and 0.9408,
although its mean PSNR increases to 35.59~dB.

Figure~\ref{fig:qualitative_reconstruction} illustrates the
reconstruction for one microscopy field. The scaled magnitude-one
initialization emphasizes boundaries and contains substantial
fine-scale variation. The final reconstruction recovers the
nuclear shapes and interior intensity patterns visible in the
reference.

Self-Diffusion nevertheless becomes much more phase consistent, with
its mean cropped phase residual decreasing from
$6.85\times10^{-3}$ to $3.62\times10^{-4}$. SelfDeblur decreases
from $1.48\times10^{-2}$ to $1.53\times10^{-3}$. Thus, lower phase
residual does not necessarily imply lower image-domain error. At the
final checkpoint, SelfDeblur gives the stronger overall image-domain
metrics and is also faster, requiring 353.4~s/image versus
541.4~s/image for Self-Diffusion.

Figure~\ref{fig:convergence} compares the iteration-to-iteration
changes of the two methods. Both neural-prior stages stabilize as the
outer iterations progress, with Self-Diffusion generally producing
smaller update magnitudes. During the final constraint-only stage,
Self-Diffusion also exhibits consistently smaller $\Delta$MSE than
SelfDeblur. SelfDeblur, in contrast, undergoes larger updates that correspond to a substantially greater improvement in image-domain quality.


\vspace{-0.25cm}
\section{Conclusion}
\vspace{-0.25cm}

We presented a two-stage iterative algorithm for phase-only image reconstruction
that combines Fourier-phase and spatial constraints with an
image-specific neural prior, which provides an initial estimate to the globally convergent constraint only refinement process. The networks
SelfDeblur and Self-Diffusion serve as alternative neural components
within this framework. Experiments on 77 microscopy images show that incorporating a neural-prior can improve reconstruction quality over constraint-only iterations at the evaluated iteration budget. Future work will address noisy or incomplete phase measurements, uncertain support, and adaptive refinement strategies.

\vfill\pagebreak

\bibliographystyle{IEEEbib}
\bibliography{strings,refs}

\end{document}